# Intrinsic Localized Modes Observed in the High Temperature Vibrational Spectrum of NaI

M. E. Manley[1], A. J. Sievers[2], J. W. Lynn[3], S. A. Kiselev[2], N. I. Agladze[2], Y. Chen[3,4], A. Llobet[5], A. Alatas[6]

[1]*Lawrence Livermore National Laboratory, Livermore, California 94551, USA*
[2]*Laboratory of Atomic and Solid State Physics, Cornell University, Ithaca, New York 14853-2501, USA*
[3]*NIST Center for Neutron Research, National Institute of Standards and Technology, Gaithersburg, Maryland 20899, USA*
[4]*Department of Materials Science and Engineering, University of Maryland, College Park, Maryland 20742, USA*
[5]*Los Alamos National Laboratory, Los Alamos, New Mexico, 87545, USA*
[6]*Argonne National Laboratory, Argonne, Illinois 60439, USA*

Abstract

Inelastic neutron measurements of the high temperature lattice excitations in NaI show that in thermal equilibrium at 555 K an intrinsic mode, localized in three dimensions, occurs at a single frequency near the center of the spectral phonon gap, polarized along [111]. At higher temperatures the intrinsic localized mode gains intensity. Higher energy inelastic neutron and x-ray scattering measurements on a room temperature NaI crystal indicate that the creation energy of the ground state of the intrinsic localized mode is 299 meV.

PACS numbers: 63.20.Pw, 63.20.Ry, 78.70.Nx, 63.20.dd

# I.   INTRODUCTION

Both classical simulations [1-4] and experiments [4-9] have shown that driving a discrete nonlinear lattice can cause some energy to spontaneously localize. Driven nonequilibrium localization experiments for a range of macroscopic and microscopic one-dimensional (1-D) lattices, from Josephson junctions to micro-mechanical systems to antiferromagnets, are summarized in recent reviews [3,4,10]. Such localized excitations now appear under a variety of names, such as discrete breather, discrete soliton, or lattice soliton. A long standing question in condensed-matter sciences and nonlinear dynamics is whether or not intrinsic 3-D localized modes (ILMs) can appear in an atomic lattice in *thermal equilibrium*. Neutron scattering measurements of phonon dispersion curves in solid bcc $^4$He [11] and in $\alpha$-U at high temperatures [12,13] have indicated extra modes, possibly attributable to ILMs, but these interpretations remain speculative since these modes remain within the extended wave spectra and realistic models of the nonlinear lattice dynamics are not available. These systems are also exceptional in that both exhibit many exotic phenomena. For example, $\alpha$-U is the only element in the Periodic Table to exhibit a charge density wave and solid bcc $^4$He is a quantum solid. The occurrence of nonlinear dynamical modes in either of these systems, while interesting, does not have broad technological implications since the underlying dynamics is related to rather unique properties.

By contrast, here we report the experimental observation of ILMs in a remarkably simple ionic crystal, NaI, at high temperatures and further show that these results are consistent with realistic molecular dynamic simulations. Neutron scattering measurements show an ILM forming in NaI in the spectral phonon gap at 0.594 of the melting temperature and above. The ILM has a polarization and 3-D localization consistent with classical predictions [14], but appears at a single frequency, not a distribution. Higher energy inelastic neutron and x-ray scattering measurements on a room temperature NaI crystal display a sharp feature at 299 meV consistent with the creation of a discrete-energy ILM ground state.

Inelastic scattering measurements were performed on a powder of pure NaI and on single crystals of NaI (+0.002Tl). This ionic crystal was chosen since molecular dynamic simulations have already demonstrated that lattice anharmonicity of the two-body potential can stabilize an ILM polarized along [111] in the phonon gap of a 3-D diatomic crystal with rigid ion NaI potentials [14], and since it had been proposed that at high temperatures such ILMs may appear in the gap, stabilized by configurational entropy, much like vacancies only with much lower activation energy [15].

## II.  EXPERIMENTS AND RESULTS

The pure NaI powder was measured on the LANSCE-PHAROS time-of-flight chopper spectrometer at Los Alamos National Laboratory from room temperature to 555 K. The incident energy was 45 meV and the detector coverage (7º to 140º) gave momentum transfers spanning several Brillouin Zones. The raw data were corrected for sample environment background, detector efficiency and the $k_i/k_f$ phase space factor. A neutron-cross-section weighted phonon density of states was extracted from the results by subtracting the elastic peak, the incoherent multiphonon scattering determined iteratively [16, 17], and dividing out the thermal occupation and Debye-Waller factors. Figure 1 shows the raw data near the gap, Fig. 1(a), and the resulting neutron-weighted phonon densities of states (DOS), Fig. 1(b). Because the inelastic scattering is dominated by the light Na atoms (4.8:1) the optic mode intensities are accentuated relative to the acoustic modes (in the true DOS the optic and acoustic regions have equal areas). In addition to the expected uniform softening of all the phonon peaks with increasing temperature there are several interesting features. First, a peak accounting for about 0.008 of the DOS develops in the gap (around 10.5 meV at 555 K). Second, the TO mode, appearing at room temperature as a strong peak around 14.5 meV, decreases in relative intensity with increasing temperature.

To obtain specific spectral information single crystals of NaI (+0.002Tl) [18] were measured using the bt7 triple-axis spectrometer at the NIST Center for Neutron Research. The spectrometer was operated with fixed final neutron energy of 14.7 meV and the crystals were mounted in a furnace with the (*hhl*) reflections in the scattering plane. Figure 2 shows the temperature dependence of energy scans in a nearly transverse geometry along [111] at the zone boundary (ZB). A correction for the two phonon component was made in the incoherent approximation [16] using a temperature corrected phonon DOS interpolated from the data in Fig. 1. At room temperature, the ZB TA, ZB LA, and ZB TO modes are all visible and have energies consistent with those reported previously [19-22]. On heating to 555 K, a peak again develops in the gap, refining to a position at 10.2 meV. The intensity of this feature in this scattering geometry indicates a strong component of polarization along [111]. At a temperature of 473 K, just below where the gap mode is evident, some intensity is lost in the ZB TO mode. The center frequencies of the ZB TA and ZB LA mode decrease with increasing temperature as expected for a thermally expanding lattice; however, the ZB TO mode frequency remains fixed near 12.8 meV between 438 and 555 K.

Using a longitudinal scattering geometry with $Q$ = (2.5, 2.5, 2.5) and $Q$ = (1.5,

1.5, 1.5) the temperature dependence of different modes was measured over a wider

temperature range, as shown in Fig. 3. With the experimental uncertainty of ±0.2 meV

the LA mode position with increasing temperature is: 9.7 meV, 300K; 8.8 meV, 555 K;

8.6 meV, 660 K; 8.4 meV, 768 K. The gap mode at 555 K now appears as a significant

shoulder at 10.3 meV on the more prominent ZB LA mode. Although forbidden in this

geometry the TO is known to appear [23]. The cause is the transverse components picked

up because of a relaxed vertical $Q$ resolution (0.34 Å$^{-1}$ at the elastic line along [0,-1,1])

[24] and a contribution from the incoherent cross section of Na. The favorable neutron

cross-section weighting on the TO relative to the LA [Fig. 1(b)] makes this effect more

noticeable. The "forbidden TO" mode and the ILM both appear broad at 660 K. By 768

K the ILM peak has gained significant intensity, as has the two phonon component. The two phonon component was again corrected in the incoherent approximation [16],

although this time by extrapolating phonon DOS from the data in Fig. 1. As a check the

768 K scan was repeated at $Q$ = (1.5, 1.5, 1.5) where the two-phonon background is much

smaller; the one-phonon intensity decreases as $Q^2$ compared to $Q^4$ for two-phonon

intensity [16]. The observed temperature dependence of the ILM strength does not appear

to follow that expected for a thermally activated process with a single activation energy

[15].

Additional measurements on dispersion properties at 555 K are summarized in intensity towards the zone center. The occurrence of a maximum at the ZB is consistent with the gap mode intensity appearing comparable to the LA at the ZB [Fig. 2, 3] and yet relatively weak in the Q averaged phonon DOS [Fig. 1(b)]. The LA mode softening with increased temperature is focused near the zone boundary, while the softening in the TO and TA branches is more uniform.

Fig. 4(a). The gap mode stands out as dispersionless across the zone, with weakening

The $Q$ dependence (structure factor) of the gap mode intensity was determined from constant-energy $Q$ scans performed across the zone near (222). Between 9 meV and 12 meV evidence of three features appear above background as shown in Fig. 4(b). Top panel: At 12 meV the $Q$ dependence is relatively flat, decreasing slightly towards the zone center, consistent with the measured TO mode dispersing slightly upward towards the zone center. Center panel: At 10 meV where the gap mode is dominant, the intensity appears smoothly spread across ~1/3 of the zone in a Gaussian shape centered at the zone boundary and fading to background near the zone center. Bottom panel: At 9 meV the gap mode intensity overlaps with the LA modes, which peak in energy near q = (0.38, 0.38, 0.38), see the LA mode dispersion in Fig. 4(a). Taken together with the dispersion curve data, it is clear that the Gaussian structure observed in the 10 meV scan is a fair representation of the gap mode structure factor. Since a Gaussian feature in reciprocal space transforms to a Gaussian feature in real space with inverse width, the local mode structure factor implies a Gaussian shaped mode about 1.2 nm across.

Since recent inelastic scattering work on uranium indicated a second much higher energy lattice excitation [13], higher energy inelastic neutron and inelastic x-ray scattering measurements have also been performed on single crystals of both NaI(+0.002Tl) at room temperature. The neutron measurements used 441 meV neutrons in the LANSCE-PHAROS time-of-flight chopper spectrometer located at Los Alamos

National Laboratory. The crystal was oriented so that the dominant direction of the $\boldsymbol{Q}$

vector was along the [111] direction in the (*hhl*) plane. The resulting spectra were

selectively summed over regions in $\boldsymbol{Q}$-space that were either near zone boundaries, where

the gap mode appears strong, or near zone centers, where the gap mode intensity is

absent, illustrated in Fig. 4(a,b). The resulting spectra summed near the zone boundaries

are shown in the top panel of Fig. 5(a). Evident is a series of three or four broad spectral

features ending with a fairly sharp excitation at 299 meV. By contrast, the sum near the

zone center displays a featureless spectrum, bottom panel in Fig. 5(a). The sharp 299

meV zone boundary feature was also observed with inelastic x-ray scattering using the

3IDC-C spectrometer at the Advanced Photon Source of Argonne National Laboratory

with incident x-ray energy of 21.657 keV. Figure 5(b) displays a weak excitation at 298

meV at the zone boundary while no excitations are observed near the zone center.

### III. DISCUSSION

The dynamical signatures of a 3-D ILM are apparent from these inelastic

scattering data on NaI. The sharp spectral feature in the gap at 555 K (Figs. 1-3) in both

powder and single crystals has a single vibrational frequency, not a distribution. The

absence of dispersion for the gap mode [Fig. 4(a)], its intensity distribution in $\boldsymbol{Q}$-space

[Fig. 4(b)], and its [111] polarization [Fig. 3], are all consistent with earlier molecular dynamics predictions in NaI of the ILM orientation and structure [14].  At the same time a number of unexpected features have been found such as the temperature independence of the ZB TO mode between 438 and 555 K [Some data is shown in Figs. 2,3]; and the softening of the LA just below the ILM near the ZB [Fig. 4(a)]. Perhaps most puzzling is the 299 meV (Fig. 5) peak observed both with INS and IXS which identifies the creation energy of the ground state of an ILM. Although requiring a large energy (~21 phonons), it is expected that these localized excitations should become visible at high temperatures because of configurational entropy [15]. The predicted ILM concentration is $\exp(-E_a/k_BT)$ = 0.0015 at 555 K. The inelastic neutron scattering intensity scales as the energy of the modes (number of occupied phonons) and since at high temperatures all normal modes have energy $k_BT$ while the ILMs are much higher at $E_a$, the apparent intensity of the ILMs is increased by $E_a/k_BT$= 6.47 giving an apparent concentration of 0.01. This is in reasonable agreement with the 0.008 number determined from the phonon DOS [Fig. 1b]. However, rather than an exponential temperature dependence, our experiments show an abrupt appearance of the ILM near 555 K followed by a broadening and then sharpening of a growing ILM peak at higher temperatures.

## IV. CONCLUSIONS

These experimental findings demonstrate that for a simple insulating ionic solid, depending on the temperature, strongly nonlinear localized excitations do appear. At elevated temperatures such vibrational gap excitations may occur with well-defined discrete energies. These excitations represent a necessary component to the phonon picture and they surely play a role in the high-temperature physical properties of crystalline solids.

### ACKNOWLEDGMENTS

Work performed under the auspices of the U.S. Department of Energy by Lawrence Livermore National Laboratory under Contract DE-AC52-07NA27344. Cornell effort supported by DOE DE-FG02-04ER46154. AJS would like to thank S. P. Love, J. B. Page and J. P. Sethna for helpful comments. The use of the Advanced Photon Source was supported by the Department of Energy, Office of Science, Office of Basic Energy Sciences, under Contract No. DE-ACOZ-06CH11357.

Figure captions

**Figure 1.** Time-of-flight inelastic neutron scattering measured on pure NaI powder at 310, 400, 500, and 555 K, summed over all detectors. **a**, Raw data near the phonon gap (555 K spectra was counted 21 hours compared to 12 hours for the others). **b**, Neutron-weighted phonon density of states (DOS). A relatively sharp feature appears near the center of the phonon gap at the highest temperature. A Gaussian peak plus a polynomial fit to the background determine the intensity of the gap feature in the DOS to be 0.008.

**Figure 2.** Inelastic-neutron-scattering spectra measured along [111] at the zone boundary of a NaI crystal in a nearly-transverse-scattering geometry (geometry shown in inset). Fits are made to multiple peaks, including the transverse acoustic (TA), longitudinal acoustic (LA), transverse optic (TO), and the ILM peak, appearing at 555 K and 10.2 meV.

**Figure 3.** Inelastic-neutron-scattering spectra measured along [111] at the zone boundary of a NaI crystal in longitudinal geometry (geometry shown in the insets). Bottom panel: The high temperature data are offset for clarity. Fits are made to the LA and the "forbidden" TO peaks at most temperatures (see text). An incoherent multiphonon scattering background has been subtracted from the data collected at 555 K and above. The gap peak at 10.3 meV at 555 K is consistent with the feature in Figs 1 and 2. At 660 K a more complex spectra is evident followed by a single mode at 768 K. Top panel: Because of a large multiphonon correction at 768 K this scan was repeated at a lower $Q$, where the multiphonon background is much smaller (this spectra is uncorrected).

**Figure 4.** (Color online) Summary of lattice dynamics measured in NaI along the [111] direction. **a,** Phonon dispersion curves at 90 K (dashed) from Ref. 19 and 555 K (solid)

from this work, including a dispersionless local mode that appears in the gap. **b,**
Constant-energy $Q$ scans at 545 K measured in a transverse geometry (geometry shown
in inset). Data from scans away from the phonons at 20 meV, shown in all scans, gives a
measure of the background. The intensity at 10 meV, middle panel, is fit to a Gaussian.
Intensities associated with the TO mode (top panel) and LA mode (bottom panel) are also
indicated, see text.

**Figure 5.** High-energy inelastic lattice response along [111] for a room temperature NaI
crystal probed using neutron and x-ray scattering. **a,** Inelastic neutron scattering (INS)
using 441 meV neutrons on the PHAROS-LANSCE spectrometer summed near zone
boundaries (top panel), and near zone centers (bottom panel). The harmonic multiphonon
contribution, which cuts off around 140 meV, was calculated in the incoherent
approximation using the measured room temperature phonon density of states (Fig. 1). **b,**
Reciprocal space sampled by the PHAROS spectrometer. Here a neutron energy loss of
299 meV is assumed and this cut in Q space is illustrated, along with the regions near and
away from the zone boundaries as applied in the detector sums. **(c)** Inelastic x-ray
scattering (IXS) collected near the 299 meV peak on the zone boundary (top panel) and
near the zone center (bottom panel).

**Figures**

Figure 1

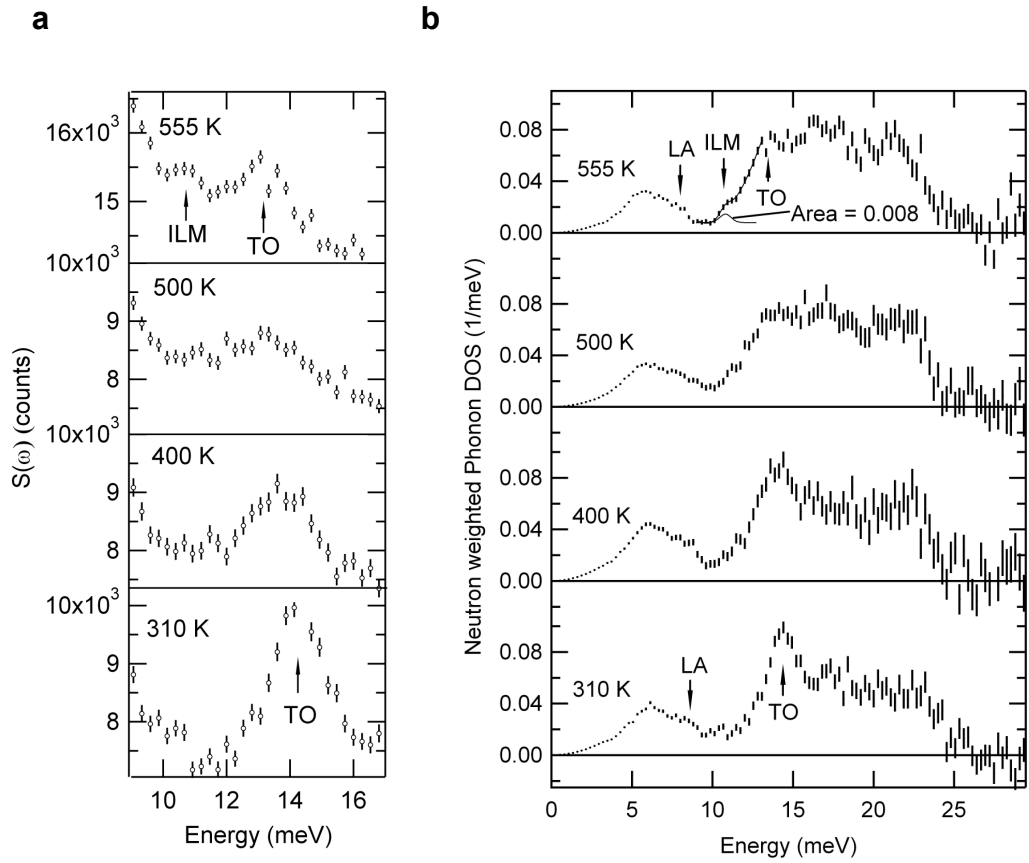

Figure 2

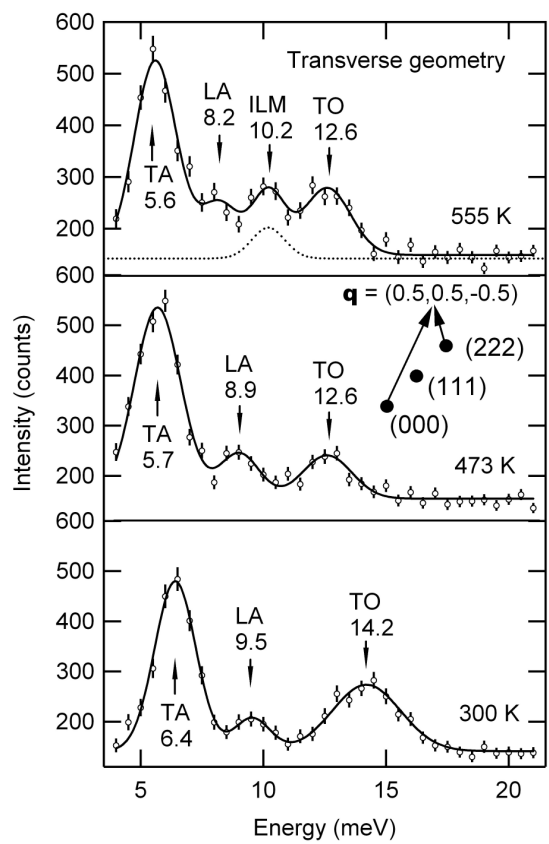

Figure 3

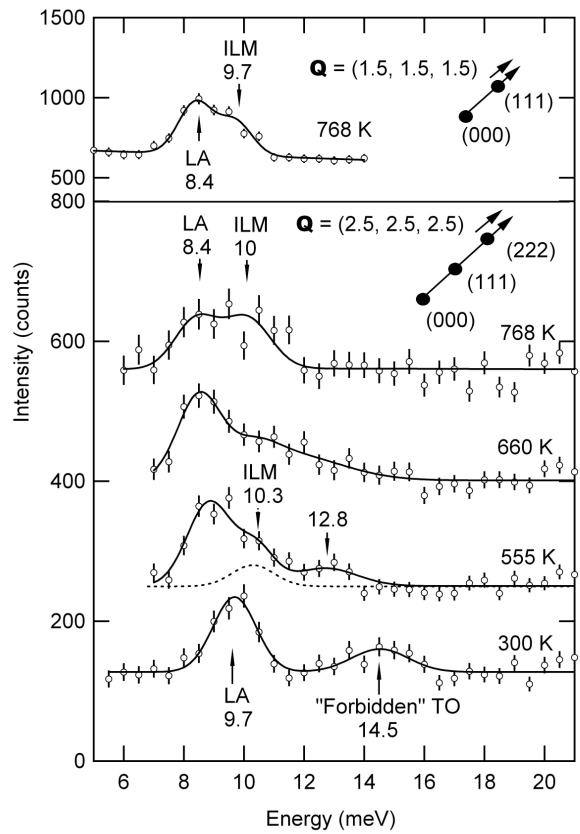

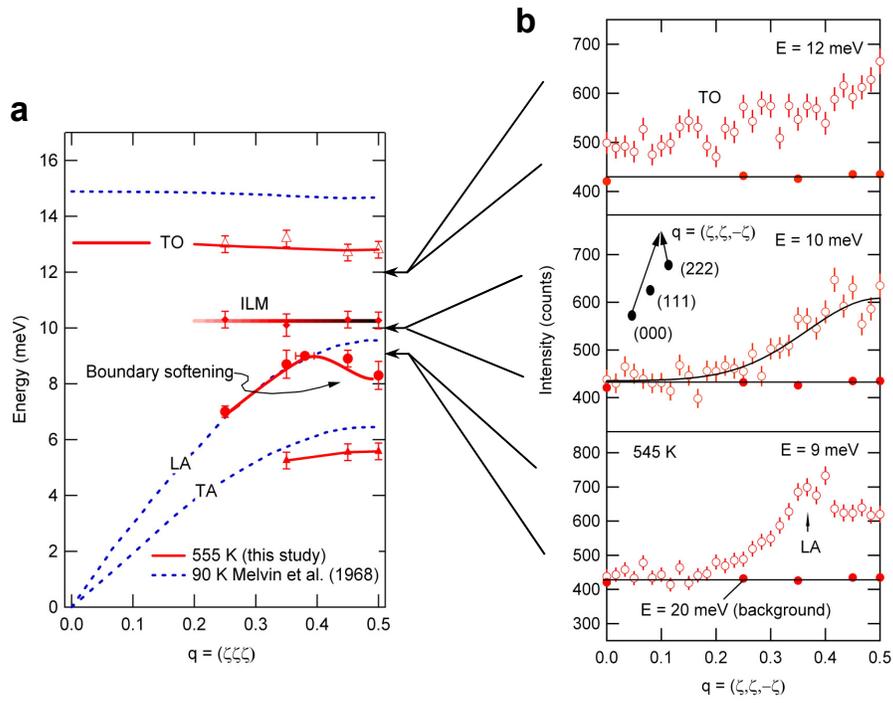

Figure 5

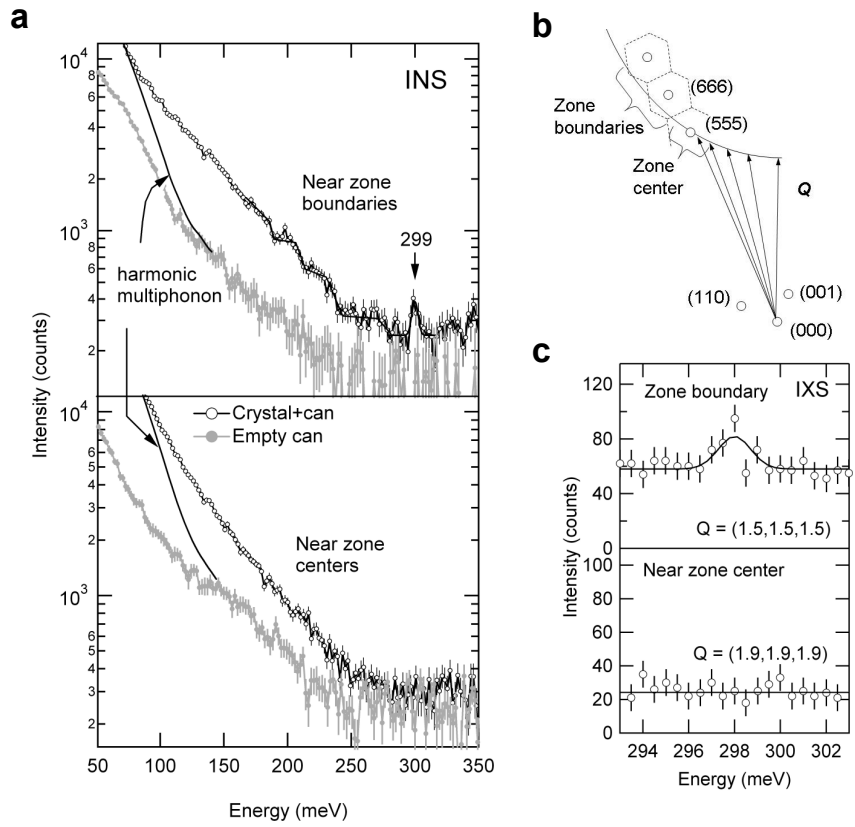